\documentclass[manuscript]{aastex}
\usepackage{emulateapj5}
\usepackage{psfig}

\newcommand{\msun}{\mbox{$\:M_{\sun}$}}

\newcommand{\expu}[3]{\mbox{\rm $#1 \times 10^{#2} \rm\:#3$}}

\begin{document}

\title{Hubble Space Telescope Observations of the Nova-Like Cataclysmic
Variable V348 Puppis\footnote{Based on observations with the NASA/ESA
Hubble Space Telescope, obtained at the Space Telescope Science Institute,
which is operated by the Association of Universities for Research in
Astronomy, Inc., under NASA contract NAS 5-2655.}}

\author{Cynthia S.\ Froning}
\email{cfroning@casa.colorado.edu}
\affil{Center for Astrophysics and Space Astronomy, \\
University of Colorado, 593 UCB, Boulder, CO, 80309}
\author{Knox S.\ Long}
\email{long@stsci.edu}
\affil{Space Telescope Science Institute, \\ 3700 San Martin Drive,
Baltimore, MD 21218}

\and

\author{Raymundo Baptista}
\email{bap@fsc.ufsc.br}
\affil{Departamento do Fisica, Universidade Federal de Santa Catarina, \\
Campus Universitario, Trindade, 88040 Florianopolis, Brazil}

\begin{abstract}

We present time series UV and optical (1150 -- 8000~\AA)
spectrophotometry of the novalike (NL) cataclysmic variable
V348~Pup through and after eclipse.  Outside of eclipse, the
spectrum is characterized by a blue continuum, strong line
emission, and a broad dip from 2000 -- 3000~\AA. The continuum
eclipse depth is roughly the same over the entire wavelength range
covered and there are no signatures of the WD eclipse in the light
curves. Model steady-state accretion disk spectra provide a poor
fit to the shape of the UV-optical spectrum outside of eclipse.
All of these properties are consistent with a disk that is
self-shielding, hiding the inner disk and the white dwarf.  In
eclipse, the emission lines and the broad dip are occulted less
than the continuum, indicating vertical extension in these
components.  The dip is consistent with the presence of a vertical
``\ion{Fe}{2}'' absorbing curtain in the system.  Archival
UV-optical spectra of the lower-inclination NL UX~UMa show a bluer
continuum and weaker line emission than in V348~Pup and no broad
dip. There is strong flickering in V348~Pup.  The flickering
spectrum shows that the flickering occurs only in the continuum.
We did not detect any periodicities associated with the rapid
variability.  The properties of V348~Pup are consistent with its
previous identification as a member of the SW~Sex sub-class of
NLs.
\end{abstract}


\section{Introduction} \label{sec_intro}

Cataclysmic variables (CVs) are interacting binary systems in which a
late-type donor star fills its Roche lobe and transfers mass to a white
dwarf (WD).  In non-magnetic CVs material is accreted onto the WD via an
accretion disk; the magnetic field strength of the WD ($\lesssim 10^{5}$~G)
is too small to significantly disrupt the accretion flow.  For an overview
of the properties of CVs, see \citet{warner1995}.

When plotted as a distribution in orbital period, very few of the known
non-magnetic CVs fall in the period range from $2.2 \leq P_{orb} \leq
2.8$~hr.  This region is known as the CV ``period gap''.  The period gap is
believed to be an evolutionary effect, described by the interrupted braking
model \citep{spruit1983,rappaport1983}.  In this picture, relatively rapid
angular momentum loss (perhaps via a magnetic wind) drives the donor stars
of longer period CVs out of thermal equilibrium, causing Roche lobe overflow
and mass transfer.  The upper edge of the period gap is created with the
sudden cessation of the rapid angular momentum loss, shrinkage of the donor
star within its Roche lobe, and a cutoff of mass transfer, which is
believed to occur when the donor star becomes fully convective.  CVs in the
period gap continue to slowly lose angular momentum via gravitational
radiation.  Eventually, the Roche lobe shrinks enough in size that it
regains contact with the surface of the donor star and mass transfer
resumes at a diminished rate.  CVs luminous enough to be observable are
thus dominated by actively accreting systems above and below the period
gap.

Nevertheless, the CV period gap is not devoid of disk-accreting CVs, and
their presence can provide an opportunity to test the interrupted braking
model and the theory of CV evolution.  One such system is V348~Pup,
discovered by \citet{tuohy1990} as the optical counterpart of a persistent
HEAO~1 X-ray source.  With an orbital period of 2.44~hr, V348~Pup lies
squarely inside the period gap.  An eclipse in its light curve and a
rotational disturbance in the radial velocities of its optical emission
lines point to the presence of an accretion disk.

V348~Pup does not show dwarf nova outbursts, so it is a novalike
(NL), a sub-class of CVs with persistently high mass accretion
rates that are nearly always found above the period gap. It has
the broad, smooth optical eclipses characteristic of eclipsing
NLs, caused by the occultation of a luminous, high accretion rate
disk. \citet{rolfe2000} showed that V348~Pup has a permanent
positive superhump from a precessing, elliptical accretion disk.
The non-axisymmetric disk morphology is driven by tidal
interactions with the donor star. \citet{rodriguez2001}
demonstrated that, based on the morphology and orbital behavior of
its line spectrum, V348~Pup can also be classified as a SW~Sex
system, a sub-type of the NLs. SW~Sex systems are characterized by
strong, single-peaked optical \ion{H}{1} and \ion{He}{1} emission
lines that are largely unocculted during eclipse, by phase lags in
the radial velocity curves of these lines relative to the motion
of the central object, and by phase-dependent absorption
components in the lines \citep{thorstensen1991}. Although SW~Sex
stars were originally thought to consist of eclipsing systems
only, several non-eclipsing systems have been discovered that
display SW~Sex type behavior \citep{martinez1999}.

V348~Pup provides an important extension of NL CVs to shorter orbital
periods. By probing the structure of the accretion disk and related
phenomena such as outflows in V348~Pup, we can better understand the
properties that govern disk accretion and how such properties depend on the
evolutionary state, mass accretion rate, and geometric size of the binary
system.  V348~Pup is eclipsing, with an inclination of 81$\arcdeg$
\citep{rolfe2000,rodriguez2001}.  As a result, it is also an excellent
candidate in which to probe the vertical structure of accretion disks by
observing the differences between eclipsed and uneclipsed components of the
disk at various wavelengths.  In order to study the accretion properties in
this unusual CV, we have obtained high-speed UV and optical
spectrophotometry of the eclipse in V348~Pup with the Faint Object
Spectrograph (FOS) on the Hubble Space Telescope (HST). The observations
and data reduction are summarized in \S~\ref{sec_obs}.  Analysis of the
data is presented in \S~\ref{sec_analysis}, and discussion and conclusions
given in \S~\ref{sec_discussion} and~\ref{sec_conc}.

\section{Observations} \label{sec_obs}

We observed V348~Pup with the HST FOS on 1996 September 11 -- 17.  There
were eight visits total.  A summary of the observations is given in
Table~\ref{tab_obs}. The visits switched between observations obtained with
the RD PRISM and observations obtained with the BL G160L grating, both
using the 1.0 aperture (0.86$\arcsec$). This combination of settings
allowed for the greatest possible wavelength coverage at high time
resolution, but at low spectral resolution. The observing strategy was the
same as one used for observations of another eclipsing NL, UX~UMa
\citep{knigge1998}. The spectra were acquired in RAPID mode, with a
spectrum obtained every 4.29~sec. The effective wavelength coverage of the
PRISM observations was 1600 -- 8000~\AA\ with a widely varying resolution
ranging from approximately 4~\AA\ at the blue end to $>$400~\AA\ at the
longest wavelengths.  The effective wavelength coverage of the G160L
grating observations was 1150 -- 2500~\AA\ at a resolution of 6.87~\AA. The
data were reduced using the standard STSDAS package routines with the
updated (POA\_CALFOS) calibration package for the FOS. In addition to the
dispersed light spectra, the zeroth-order undispersed light was recorded in
observations with the G160L grating.  For calibrating the undispersed data,
we used the rough (to within 50\%) absolute flux calibration given in
\citet{eracleous1994}.

Due to non-repeatability in the placement of the filter grating
wheel (FGW), the absolute wavelength solution for FOS spectra in
an observation can be off by as much as 1~diode. At the longest
wavelengths in the PRISM setting, this can translate to an error
of hundreds of \AA.  We did not obtain wavelength calibration
spectra during our observations so we do not have the means to
recalibrate the wavelength solution.  Fortunately, the effect of
this error is to shift the entire spectrum by the same amount with
every spectrum in an observation having the same shift (based on
the placement of the FGW at the start of the observation).  As a
result, it is the absolute wavelength solution for each
observation only that is uncertain, not the relative wavelength
solutions of each spectrum.  The FOS paper products from the
observations confirm that our observations were normal, and that
other potential sources of wavelength uncertainty (such as
mis-centering of the target in the aperture) did not occur.

We see absolute wavelength offsets in several of our visits of
order 1~diode, so for the PRISM spectra we determined the
wavelength shift for each visit using the position of the
\ion{Mg}{2} $\lambda2800$~\AA\ feature and corrected the
wavelength solution accordingly.  We also recalculated the flux
calibration solution to ensure that the correct wavelength and
sensitivity were matched for each pixel.  For the G160L grating
spectra, there were no obvious wavelength offsets between the
various observations but, based on the wavelengths of various
spectral features, all spectra were shifted by about 1 diode, or
6.87~\AA, to the red.  We have corrected this offset in the
analysis that follows.

The orbital phases for the spectra were determined using the
ephemeris of \citet{baptista1996}\footnote{A revised ephemeris has
been published for V348~Pup \citep{rolfe2000}. The two orbital
solutions agree within their uncertainties, so we did not rebin
the data.}.  
Three of the visits in each dispersion configuration
were obtained during eclipse (Runs 1--4, 7--8 in
Table~\ref{tab_obs}), and one observation each in the prism and
grating configurations was obtained after primary eclipse (Runs 5
\& 6).

\section{Analysis} \label{sec_analysis}

\subsection{Spectra and Light Curves Through Eclipse}

\begin{figure}
\psfig{file=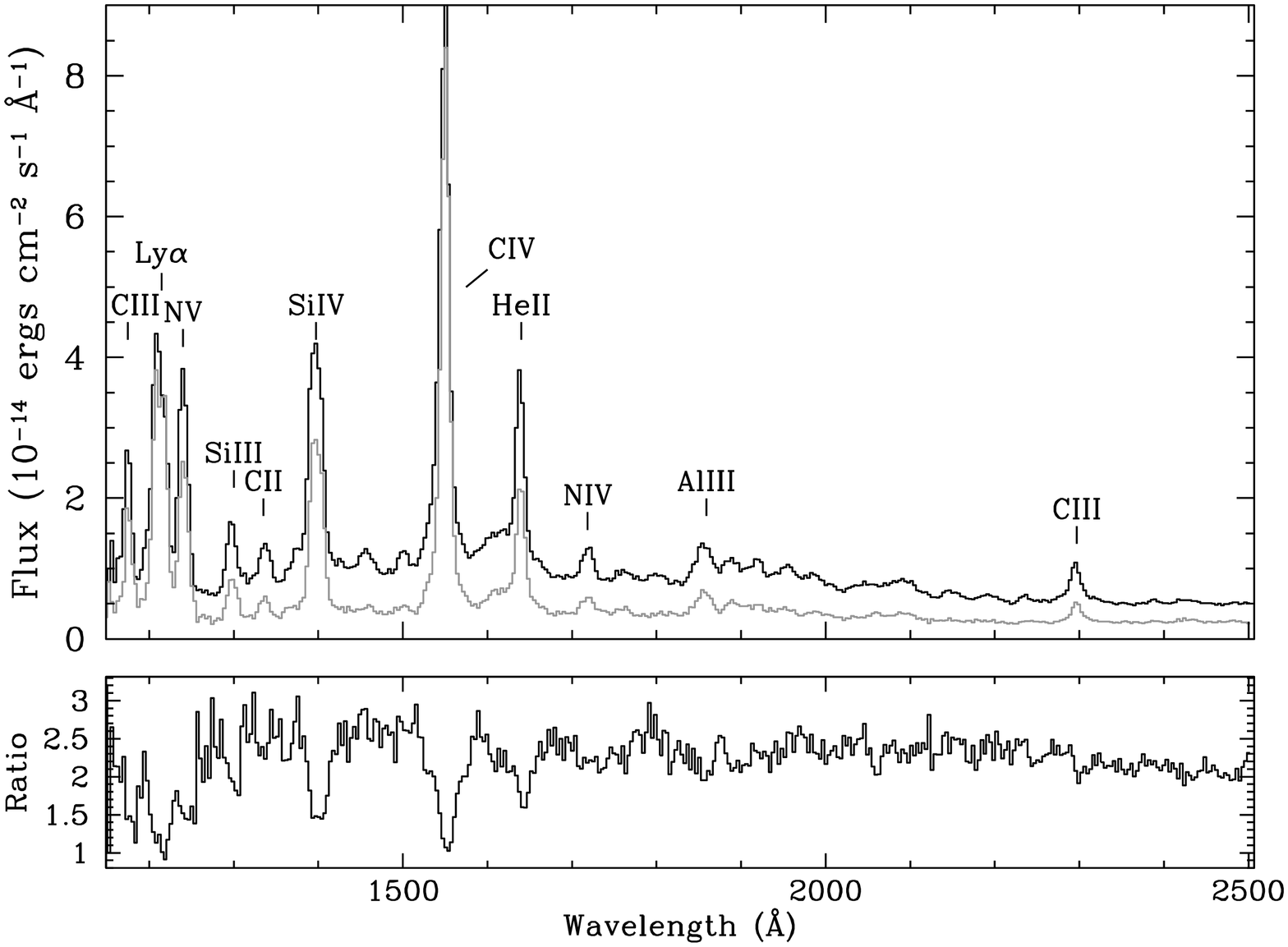,angle=-90,width=7in}
\figcaption[grat.ps]{The out of eclipse and mid-eclipse G160L
spectra of V348~Pup.  The time-averaged spectrum for phases
outside of eclipse ($\phi < 0.92$ and $\phi > 0.08$~cycles) is
shown in the upper panel in black.  The mid-eclipse spectrum
($-0.035 < \phi < 0.035$~cycles) is shown in the upper panel in
gray.  The lower panel shows the ratio of the out of eclipse to
mid-eclipse spectra. Prominent spectral features are labeled.
\label{fig_grat}}
\end{figure}

\begin{figure}
\psfig{file=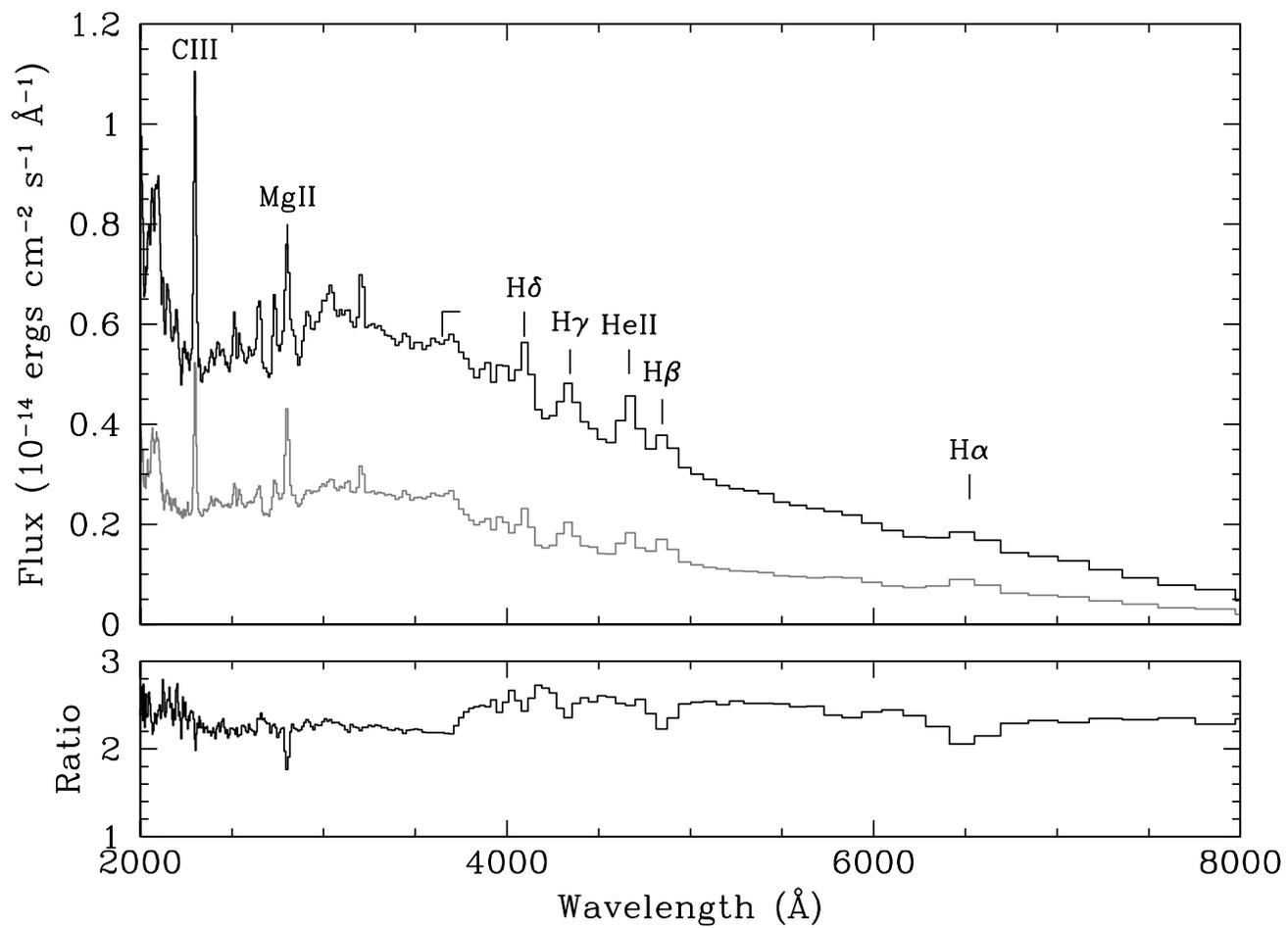,angle=-90,width=7in}
\figcaption[prism.ps]{The out of eclipse and mid-eclipse spectra
and the ratio of the same from the PRISM observations of V348~Pup.
\label{fig_prism}}
\end{figure}

Figure~\ref{fig_grat} shows the combined out-of-eclipse and
mid-eclipse spectra from the grating observations, while
Figure~\ref{fig_prism} shows the same for the prism observations.
There were no gross changes in the flux or shape of the spectra
between observations, so all data were combined together. Here and
elsewhere in this paper, out-of-eclipse phases are defined as
$\phi < 0.92$ and $\phi > 0.08$, and include the observation in
each setting acquired post-eclipse.  Mid-eclipse is defined as the
duration of total eclipse of the WD, 0.07~cycles (assuming $q$ =
0.31 and $i$ = 81$\arcdeg$; Rolfe et al.\ 2000, Rodr\'{\i}guez-Gil
et al.\ 2001), so that mid-eclipse runs from $\phi$ = 0.965 --
0.035. Also shown in the bottom panel of each figure is the ratio
of the out of eclipse to mid-eclipse spectra.

Outside of eclipse, the broadband UV-optical spectrum rises to the
blue, although there is a broad trough in the spectrum from 2000
-- 3000~\AA. The spectra show strong emission lines of \ion{H}{1}
(Balmer series and Ly$\alpha$); \ion{He}{2}; resonance transitions
of \ion{C}{2}, \ion{C}{4}, \ion{N}{5}, \ion{Mg}{2}, \ion{Al}{3},
and \ion{Si}{4}; and excited state transitions of \ion{C}{3},
\ion{N}{4}, and \ion{Si}{3}. The Balmer jump is shallow or absent.
The resonance doublet of \ion{C}{4} $\lambda\lambda$1548,1552~\AA\
is the strongest line in the spectrum, with a combined EW of
$-170$~\AA.  The FWHM of the UV lines range from 2500 --
4000~km~s$^{-1}$. (The FHWM measurement assumed Gaussian profiles
for the single lines and double Gaussian profiles for the
doublets, where the FWHM was fixed to the same value for both
components.)

The depth of the continuum eclipse is similar over the entire
UV-optical wavelength range. At most continuum wavelengths, 55\%
-- 60\% of the out-of-eclipse flux is occulted. The eclipse is
shallower in the broad spectral trough from 2000 -- 3000~\AA, with
only 45\% of the out of eclipse flux occulted at 2500~\AA.  The
eclipse depth remains around 50\% to the red of this trough until
the Balmer limit, where the eclipse depth rises sharply back to
60\% of the out of eclipse flux occulted.  As a result, the
mid-eclipse spectrum has an even weaker Balmer jump than the out
of eclipse spectrum.  The 2000 -- 3000~\AA\ trough is also more
shallow relative to the rest of the spectrum at mid-eclipse than
outside of eclipse.

All of the emission lines seen in the out-of-eclipse spectrum are
also present at mid-eclipse, indicating that they are not fully
occulted.  The degree to which the lines are eclipsed varies. The
strongest lines, Ly$\alpha$ and \ion{C}{4}, are essentially
unobscured near their line cores. Ly$\alpha$ is most likely
contaminated by variable airglow emission, so we will not discuss
its properties further, but the shallow eclipse in \ion{C}{4} is
intrinsic to V348~Pup. \ion{C}{3} $\lambda$1175~\AA\ also has a
very shallow eclipse. In the rest of the emission lines, 20\% --
55\% of the out of eclipse flux near line center is occulted. For
all of the emission lines, the eclipse is weaker in the line than
in the adjacent continuum.

\begin{figure}
\psfig{file=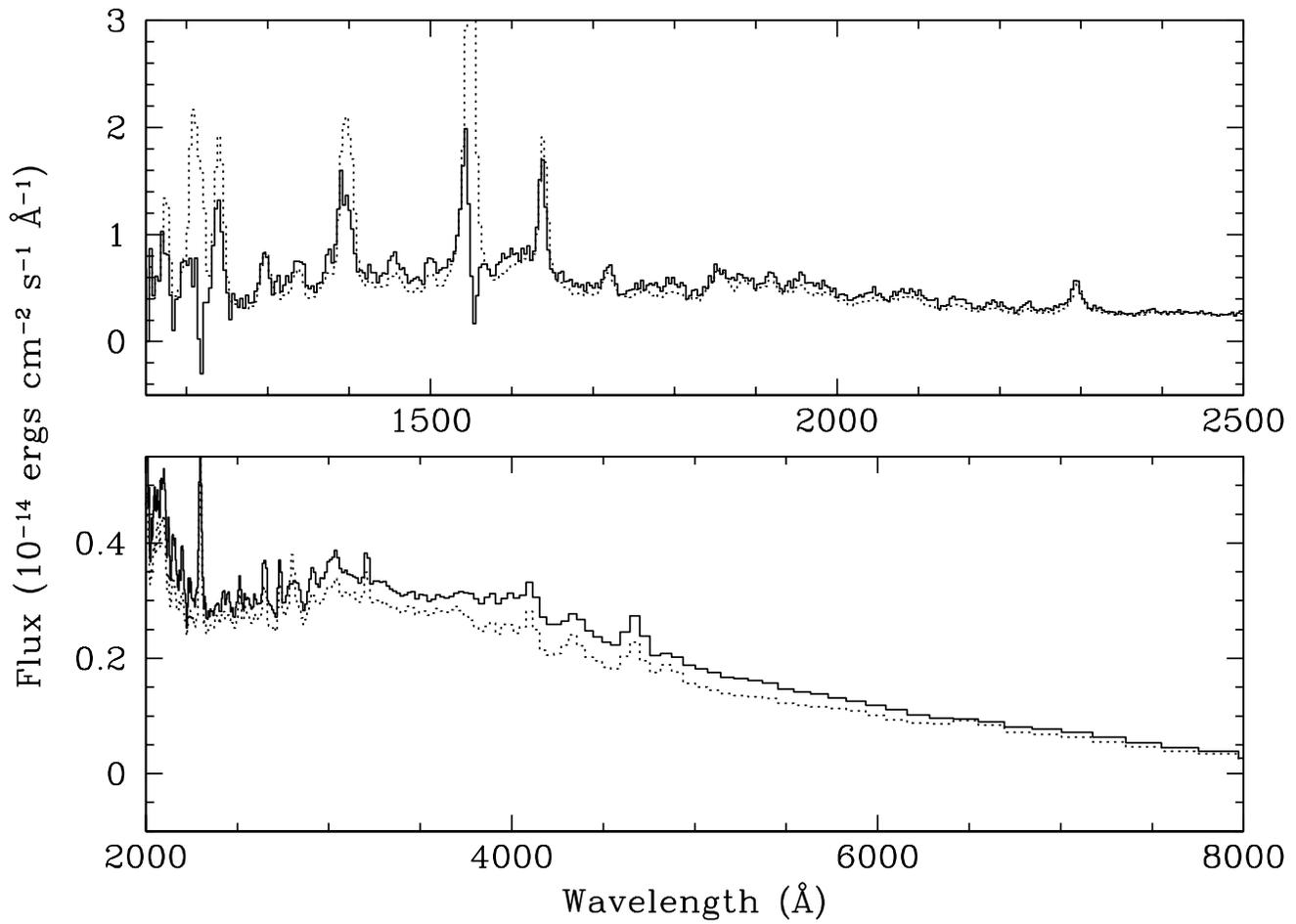,angle=-90,width=7in}
\figcaption[eclipsed.ps]{The spectrum of the eclipsed light from the G160L
(upper panel) and PRISM (lower panel) observations.  The eclipsed light
spectrum, shown as the solid line in each panel, is the difference spectrum
between the out-of-eclipse and mid-eclipse spectra.  Also shown for
reference as the dotted line in each panel is the out-of-eclipse spectrum,
scaled down by a factor of two. \label{fig_eclipsed}}a
\end{figure}

For most of the UV and optical emission lines, the line profile is
unchanged in eclipse.  Interestingly, however, the eclipse profile
is not symmetric about the line center in \ion{C}{4} and
\ion{C}{3} $\lambda$1175~\AA: the eclipse depth is smaller to the
red of line center than to the blue.  A similar but less
conspicuous asymmetry in the eclipse minimum with respect to line
center can be seen in \ion{Al}{3} $\lambda$1855~\AA\ and
\ion{C}{3} $\lambda$2300~\AA. Figure~\ref{fig_eclipsed} shows the
spectrum of the eclipsed light (that is, the difference spectrum
between the out-of-eclipse and mid-eclipse spectra).  For most of
the lines, the eclipsed light spectrum is much the same in line
profile and morphology as the out-of-eclipse spectrum, but for
\ion{C}{4} and \ion{C}{3} $\lambda$1175~\AA, the line profile in
eclipsed light has a blue emission component and a red trough that
actually falls below the local continuum into absorption. The peak
in \ion{Si}{4} $\lambda\lambda$1393,1402~\AA\ is also blueshifted
in eclipse relative to its peak out of eclipse, but there is no
absorption component in this line in the eclipsed light spectrum.

\begin{figure}
\plotone{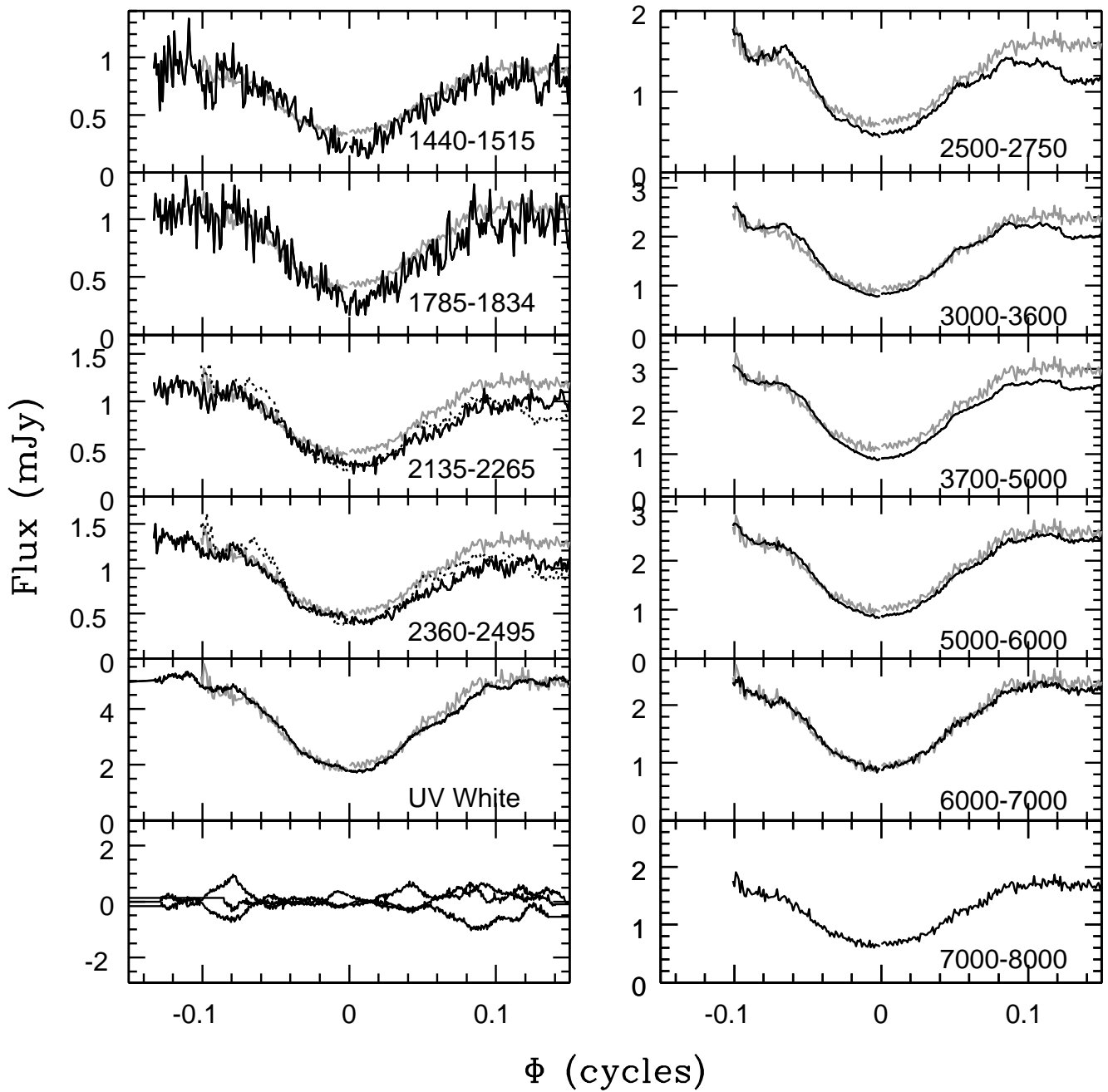}
\figcaption[v348puplc.ps]{Continuum light curves from the G160L
and PRISM eclipse observations of V348~Pup, shown in black.  The
three eclipse observations from each setting have been averaged
and binned to 2000 points per orbit, or 0.0005~cycles orbital
phase resolution. The wavelength ranges for each light curve are
given in each panel.  In panels where two curves are seen, the
spectral range was covered by both observation settings, and the
PRISM observations are shown as the dotted line.  The ``UV White''
light curve near in the lower left corner is the eclipse in
undispersed UV light from the G160L observations ($\lambda_{eff}$
= 3400~\AA). The lowest left panel shows the residuals of the
three G160L observations in undispersed light with respect to the
mean light curve above it. Also repeated for comparison in gray in
each panel is the 7000 -- 8000~\AA\ light curve, scaled to the
pre-eclipse flux of the light curve in the panel. \label{fig_lcs}}
\end{figure}

In Figure~\ref{fig_lcs}, we show continuum eclipse light curves in
several bands across the UV-optical range.  The light curves are
binned (2000 points per binary orbit) averages of the three
eclipse observations obtained in each observation setting.  Also
shown in the lower left panel are the residuals from the three
G160L eclipse observations in undispersed light with respect to
the average light curve.  The residuals show that flickering is
strong in V348~Pup.  The behavior and properties of the flickering
are discussed in \S\ref{sec_flickering}, but it should be noted
here that the strength of the flickering confuses the eclipse
profile. We have averaged the three eclipse observations in each
observation setting, but three observations are not always
sufficient to separate orbital variability effects from secular
fluctuations.

The continuum eclipse light curves are quite similar over the
broad wavelength coverage of our observations.  All of the
continuum eclipses are V-shaped, particularly in contrast to the
smooth, U-shaped eclipses in UX~UMa (see \S\ref{sec_uxuma}).  As
noted before, there is little change in the depth of the eclipse
from UV to optical. There is also no sign of a change in the
eclipse width over the wavelength range.  While the red light
curves ($>5000$~\AA) show the same pre- and post-eclipse flux
levels, the post-eclipse flux does not recover to the pre-eclipse
levels in the UV and blue optical.  Both the modest change in
eclipse shape with wavelength and the V-shaped eclipses point to
the eclipse of a source that has a relatively flat radial
temperature distribution.

Finally, there is no clear signature of the eclipse of the WD or,
indeed, any other ingress or egress features in the light curves.
A 15,000~K -- 30,000~K WD is most observable at short wavelengths,
where its fractional contribution to the observed flux is
maximized. There is no sign of the WD in the UV light curves,
however.  An examination of the residuals of the individual light
curves with respect to the mean curve in the lower left panel of
Figure~\ref{fig_lcs} shows that what appear to be possible
ingress/egress features in the UV light curves are actually
associated with flickers and are not repeated in every eclipse
observation.

\subsubsection{A Comparison of the Spectra and Light Curves of
V348~Pup, UX~UMa, and DW~UMa} \label{sec_uxuma}

\begin{figure}
\psfig{file=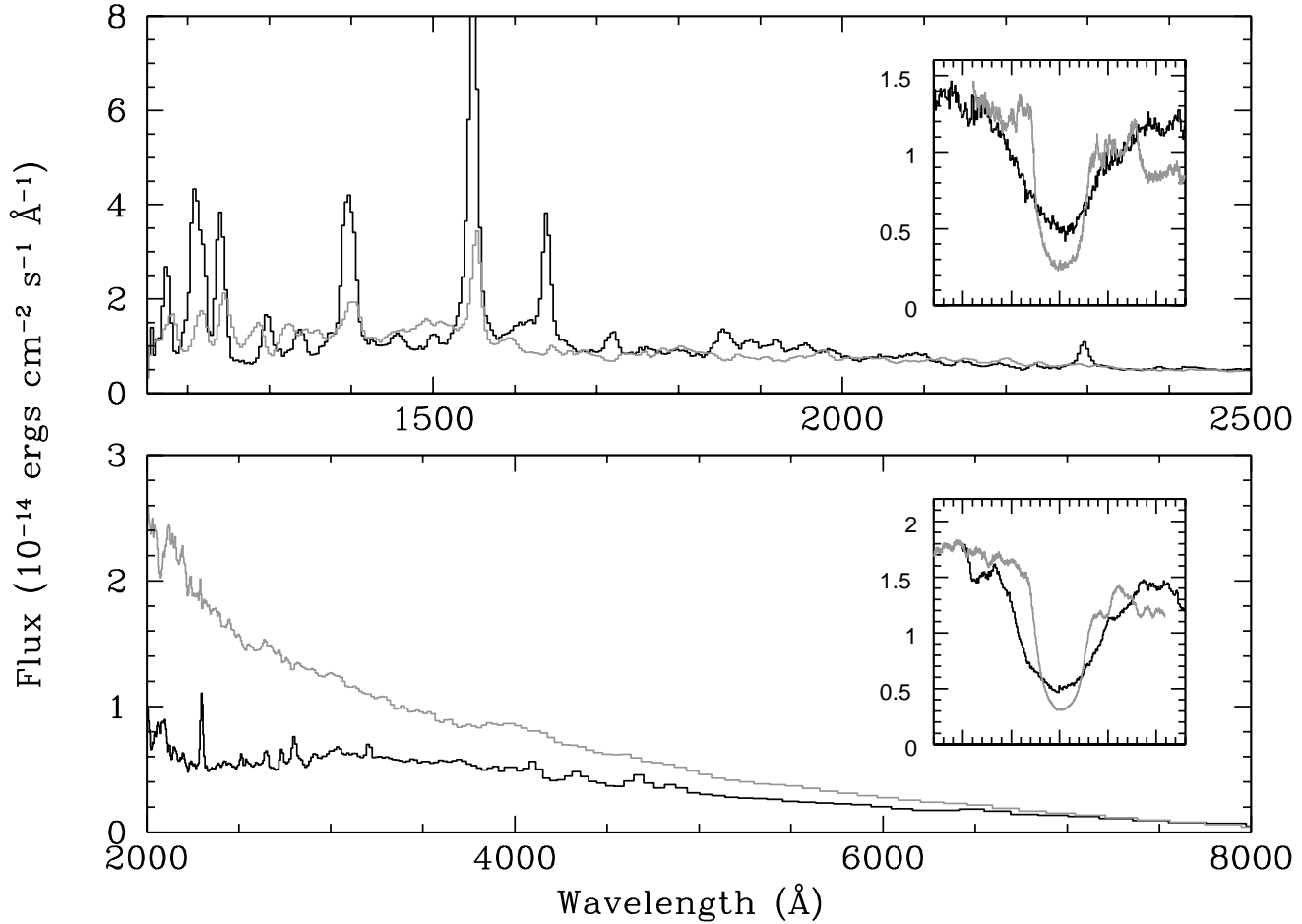,angle=-90,width=7in}
\figcaption[uxuma.ps]{A comparison of the spectra and eclipse
light curves of V348~Pup and the eclipsing NL UX~UMa. Shown in
black are the out of eclipse spectra of V348~Pup from the G160L
(upper panel) and PRISM (lower panel) observations.  Overplotted
in gray are the out of eclipse spectra from the 1994 FOS
observations of UX~UMa.  The G160L spectrum has been divided by 12
so that its flux is equal to that of V348~Pup at 2500~\AA. The
PRISM spectrum has been divided by 6 so that its flux is equal to
that of V348~Pup at 8000~\AA. (The UX~UMa G160L and PRISM
observations were obtained in two different epochs with different
fluxes, and therefore require separate normalizations.) Wavelength
offsets between the emission lines in each target may result from
uncertainties in the absolute wavelength calibrations of both
observations and should be disregarded. Shown in the inset panels
are the eclipse light curves of V348~Pup and UX~UMa. In both
plots, the y-axis units are in $F_{\nu}$ (mJy) and the x-axis
units are in orbital phase from -0.13 -- 0.13 cycles. The upper
panel shows in black the 1150 -- 2500~\AA\ average light curve of
the V348~Pup G160L observations. Shown in gray is the 1150 --
2500~\AA\ light curve of UX~UMa from its G160L observations.  The
UX~UMa light curve has been divided by 12.  The lower panel inset
shows the 2000 -- 8000~\AA\ average light curve of V348~Pup in
black, and the 2000 -- 8000~\AA\ average light curve of UX~UMa in
gray. The UX~UMa light curve has been divided by 50.  For
simplicity, we do not average and bin the two eclipse observations
of UX~UMa in each setting, but just show the unbinned light curve
of the first eclipse observation in each setting.
\label{fig_uxuma}}
\end{figure}

Since the eclipsing NL UX~UMa, a system above the period gap, was observed
with the same setup as our observations, it is instructive to compare the
two systems. Detailed analysis of the UX~UMa observations can be found in
\citet{knigge1998} and \citet{baptista1998}. UX~UMa is an eclipsing NL with
an orbital period of 4.72~hr and an inclination of 71$\arcdeg$
\citep{baptista1995}. Figure~\ref{fig_uxuma} shows the out of eclipse G160L
and PRISM spectra of V348~Pup and UX~UMa, and broadband light curves of
both objects.  UX~UMa is brighter than V348~Pup, so its spectra and light
curves have been scaled down to the fluxes of V348~Pup for comparison.

The rich emission line spectrum of V348~Pup in the UV is similar
to that of UX~UMa. The emission lines are not as strong in UX~UMa
than in V348~Pup, however, and \ion{He}{2} emission at 1640~\AA\
and 4686~\AA\ is significantly weaker.  The Balmer jump is more
prominently in absorption in UX~UMa than in V348~Pup. The
continuum is significantly bluer in UX~UMa: when the two spectra
are fixed to the same flux at 2500~\AA, UX~UMa is 20\% brighter
than V348~Pup at 1500~\AA\ and 25\% brighter at 1275~\AA. In the
PRISM spectra, UX~UMa exceeds V348~Pup in flux by 60\% at
4000~\AA\ and 240\% at 2000~\AA\ (when fixed to the same flux at
8000~\AA). Interestingly, UX~UMa does not show the broad 2000 --
3000~\AA\ spectral dip that is seen in V348~Pup.

A comparison of the light curves indicates that the eclipse is wider in
V348~Pup than in UX~UMa. The wider eclipse is due to the combination of a
higher inclination (81$\arcdeg$ vs.\ 71$\arcdeg$ in UX~UMa; Baptista et
al.\ 1995) and a larger disk size in V348~Pup. By modeling the superhump in
V348~Pup, \citet{rolfe2000} determined that the disk emission extends out
to 80 -- 90\% of the primary Roche lobe radius, significantly farther than
the outer disk emitting radius in UX~UMa ($\simeq$0.6~R$_{L_{1}}$; Baptista
et al.\ 1998).  UX~UMa has a smooth, U-shaped eclipse with a steep ingress
and egress, while the eclipse in V348~Pup is more V-shaped, with a gradual
ingress and egress and relatively short duration of eclipse
minimum. Finally, UX~UMa has a stronger pre- versus post-eclipse flux
variation than in V348~Pup, although this can change dramatically from
eclipse to eclipse in UX~UMa, with the post-eclipse flux often exceeding
the pre-eclipse level \citep{smak1994}.

\begin{figure}
\psfig{file=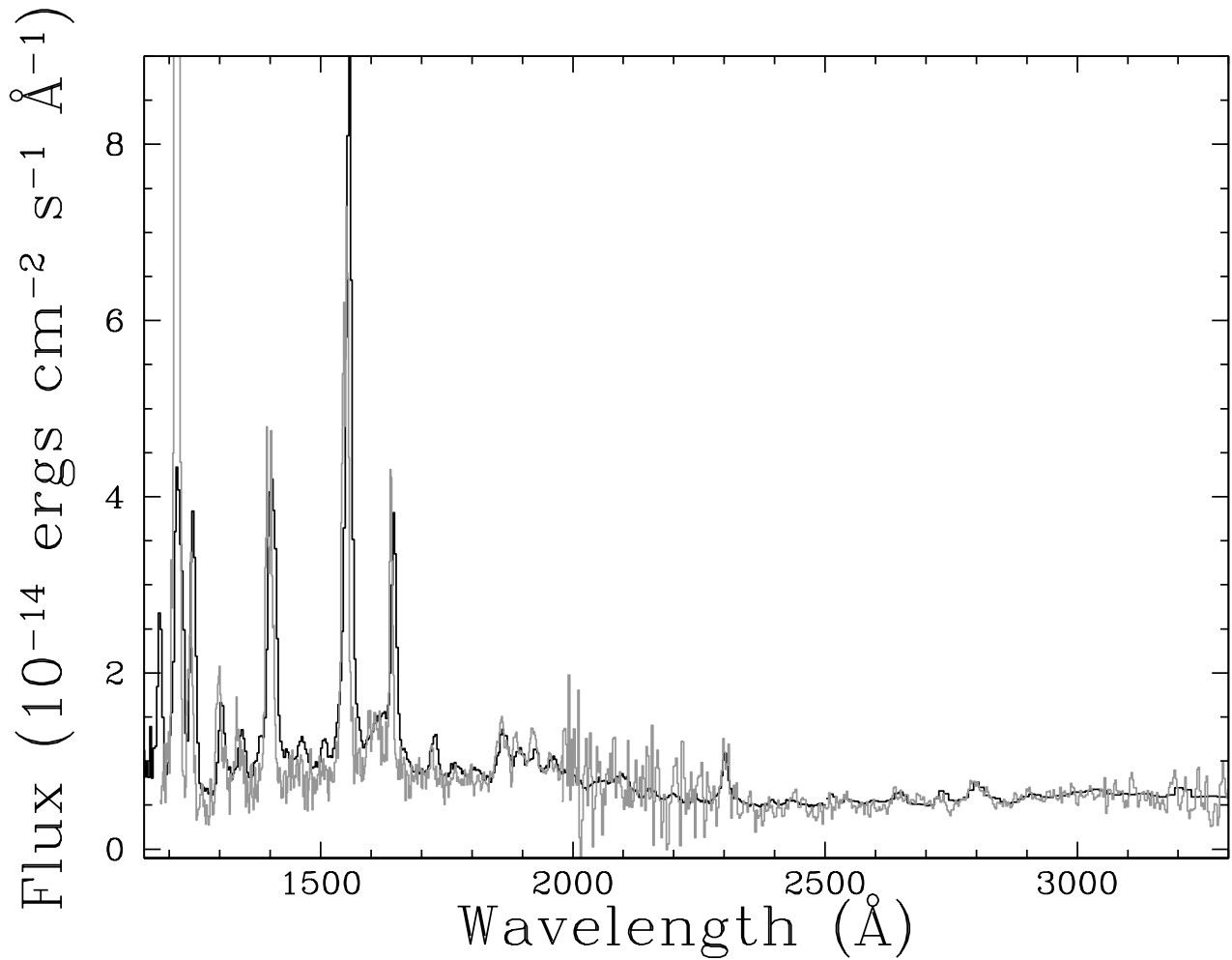,angle=-90,width=7in}
\figcaption[dwuma.ps]{A comparison of the UV spectra of V348~Pup
and the eclipsing NL DW~UMa.  Shown in black are the out of
eclipse G160L and PRISM spectra of V348~Pup (the switch between
spectra occurs at 2500~\AA).  Overplotted in gray are the
time-averaged SWP and LWP IUE spectra of DW~UMa from observations
in 1985 and 1987.  The DW~UMa spectra have been divided by a
factor of 2. \label{fig_dwuma}}
\end{figure}

The eclipsing NL DW~UMa is also interesting to compare to V348~Pup because
the two systems have the same inclination ($i = 82\arcdeg\pm4\arcdeg$ for
DW~UMa; Araujo-Betancor et al.\ 2003).  DW~UMa has an orbital period of
3.28~hr. Although there are no broadband UV-optical observations of DW~UMa
in its normal state, it was observed in 1985 and 1987 by IUE. The IUE
observations are discussed in greater detail in \citet{szkody1987}.  Here
we compare the out of eclipse spectra of V348~Pup and DW~UMa, both shown in
Figure~\ref{fig_dwuma} (the DW~UMa spectrum has been scaled down by a
factor of 2).  In the shape of the UV continuum and the appearance and
strength of the emission lines, the two sources have UV spectra that are
virtually identical.  The drop in flux longward of 2000~\AA\ suggests that
the broad 2000 -- 3000~\AA\ dip seen in V348~Pup is also present in DW~UMa.

\subsection{Pre- Versus Post-Eclipse Spectra} \label{sec_prepost}

There are indications that the spectrum of V348~Pup is not the
same in the phases before and after eclipse:  the UV and blue
optical fluxes are smaller after eclipse than before, while the
red optical flux is the same before and after eclipse. As a
result, we undertook a comparison of the pre- and post-eclipse
spectra in V348~Pup. We combined all spectra obtained from orbital
phases 0.88 -- 0.92 for pre-eclipse and all spectra from phases
0.08 -- 0.12 for the (immediate) post-eclipse spectrum. We
combined all spectra within the defined orbital phase ranges from
the three eclipse observations in each observation setting to
create mean pre- and post-eclipse spectra.  Two of the three
eclipses were well covered before and after eclipse in each
observation setting, while the pre-eclipse phases were largely cut
off in one eclipse observation in each setting.

We are somewhat limited in our ability to draw conclusions concerning the
pre- and post-eclipse behavior of the spectrum, however, because the
relative fluxes before and after eclipse changed from observation to
observation.  In one of the G160L eclipse observations, the 1500~\AA\ flux
dropped 30\% between the mean pre- and post-eclipse levels.  In the other
observation with pre-eclipse coverage, however, there is no difference in
the continuum before and after eclipse. Similarly, of the two PRISM
observations spanning the full eclipse, one shows a 20\% drop in flux at
3000~\AA\ before and after eclipse and one shows no drop.  We therefore
discuss below those features that appear in multiple eclipses.

\begin{figure}
\psfig{file=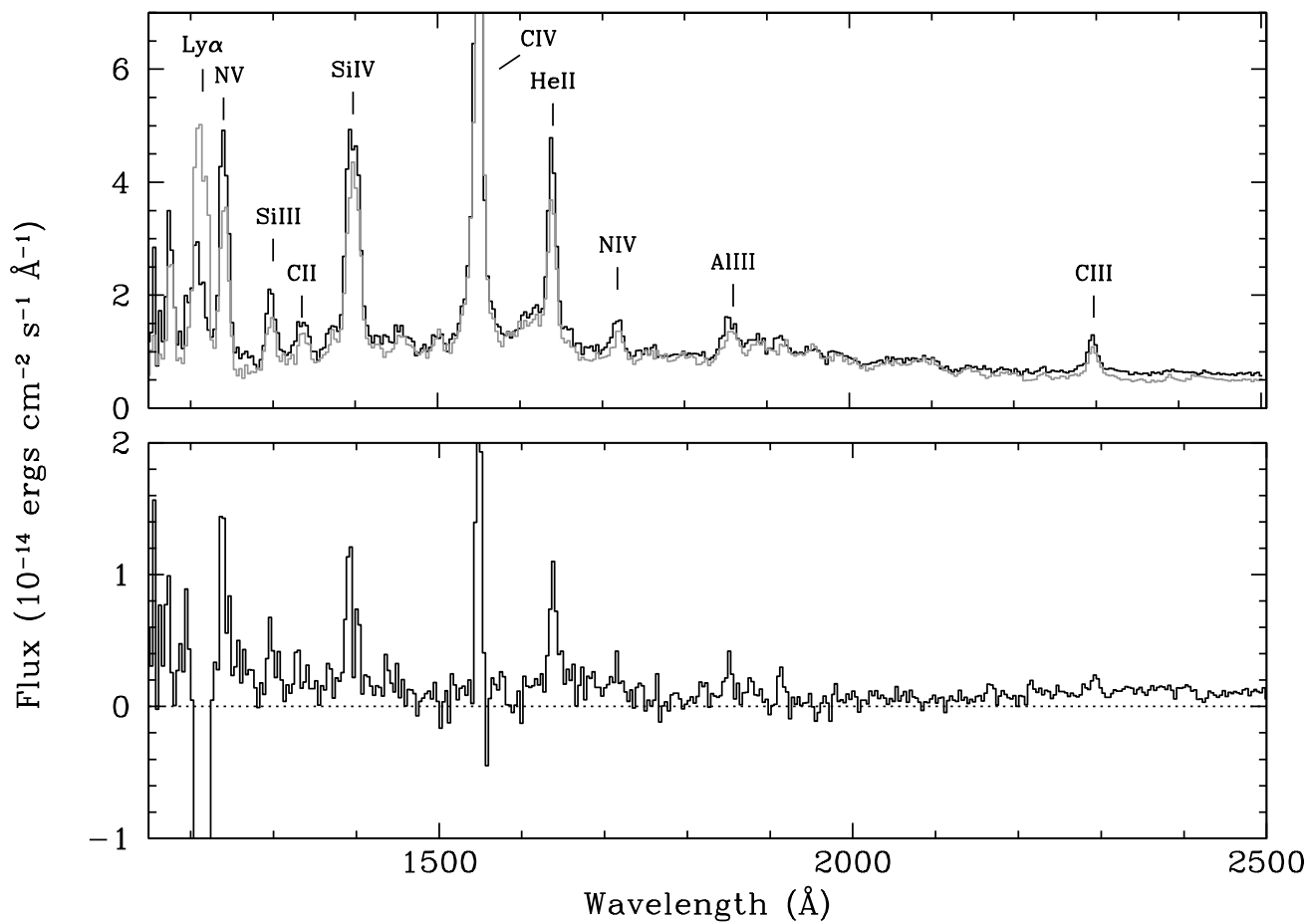,angle=-90,width=7in}
\figcaption[prepostgrat.ps]{Mean spectra before and after eclipse,
and the difference spectrum between them.  The upper panel shows
the pre-eclipse G160L spectrum in black and the post-eclipse
spectrum in gray.  The pre-eclipse spectrum is an average of all
spectra acquired from orbital phases 0.88 -- 0.92.  The
post-eclipse spectrum is an average of 0.08 -- 0.12 cycles
spectra. Shown in the lower panel is the difference spectrum
between pre- and post-eclipse. \label{fig_prepostgrat}}
\end{figure}

\begin{figure}
\psfig{file=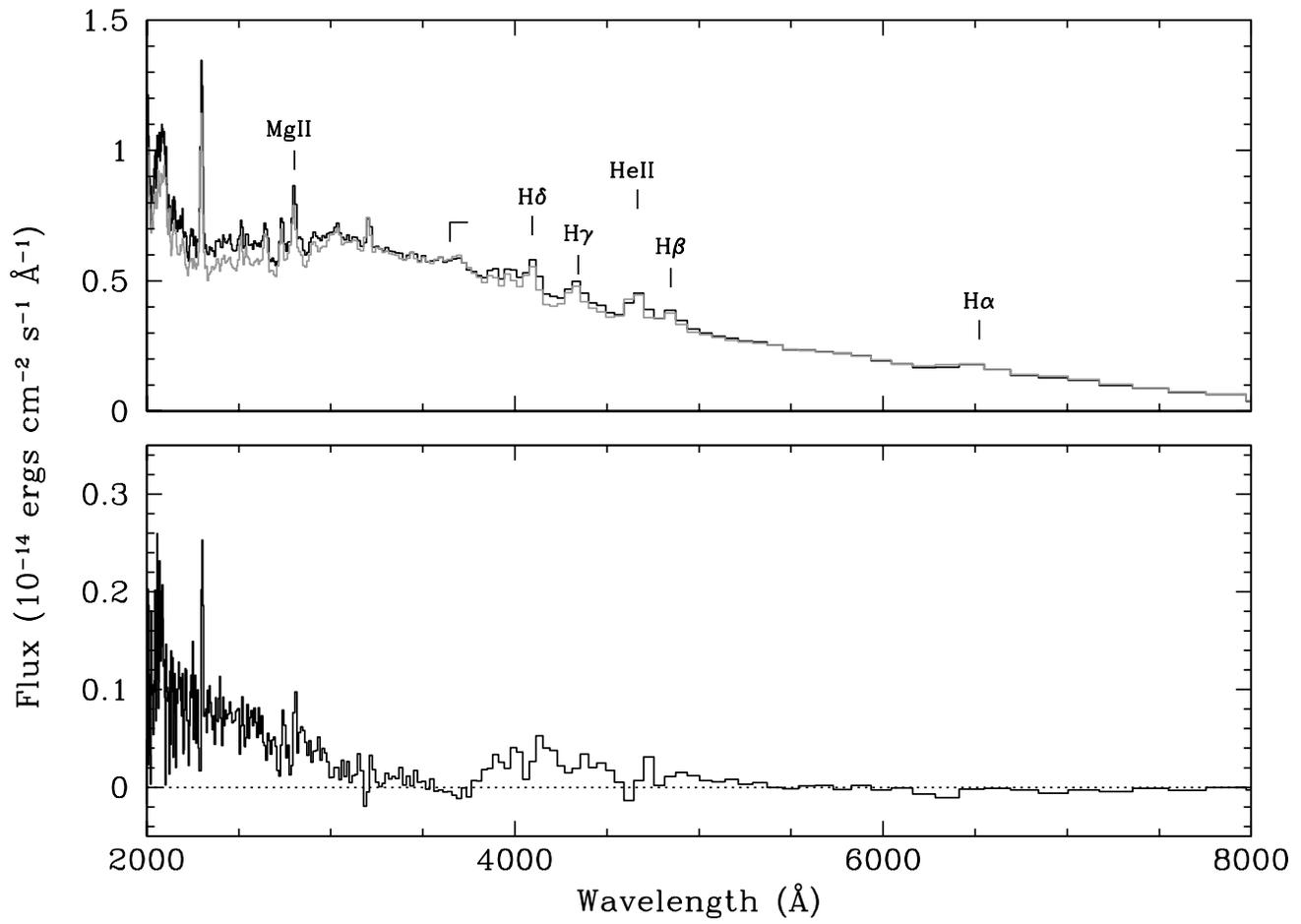,angle=-90,width=7in}
\figcaption[prepostprism.ps]{Mean spectra before and after eclipse, and the
difference spectrum between them for the PRISM observations.
\label{fig_prepostprism}}
\end{figure}

The mean pre- and post-eclipse spectra, and the difference spectra, are
shown in Figures~\ref{fig_prepostgrat} and~\ref{fig_prepostprism}.  As
suggested by the eclipse light curves, there is a difference in the
spectrum before and after the eclipse.  The continuum red of the Balmer
limit is roughly the same before and after eclipse, but the blue emission
does not recover its pre-eclipse level. Most of the UV and optical emission
lines are also weaker after eclipse than before.  The spectral dip from
2000 -- 3000~\AA\ is deeper (i.e., lower in flux) after eclipse than
before. The resulting difference spectrum between pre- and post-eclipse
shows a blue continuum and line emission.

Interestingly, the two observations obtained after the eclipse (Runs 5 \&
6) also show a steady decline in the observed flux, extending to at least
phase 0.3. The drop is 20\% in the PRISM observation and 45\% in the G160L
observation from phase 0.1 -- 0.3.  This suggests that there is an orbital
phase-related drop in the flux starting before the eclipse and extending
well after eclipse end, but with our limited data, little more can be said
on this point.

There is more than one possible source for the variable emission across the
eclipse. The bright spot, where the mass accretion stream impacts the edge
of the disk, has an orbital phase behavior consistent with a difference in
pre- and post-eclipse flux.  If this component is the bright spot, its
properties are in contrast to those of the bright spot in UX~UMa, which
showed lines in absorption, the opposite of what we see here.  The
permanent superhump could also cause a slowly-varying fluctuation across
and beyond the eclipse (see Rolfe et al.\ 2000, Fig.\ 7).  Unfortunately,
the superhump phasing is very unstable, and we do not have enough data to
determine the superhump phase at the time of our observations.

\subsection{Light Curve Flickering} \label{sec_flickering}

The light curves in V348~Pup show frequent, large amplitude secular
variability, or flickering, especially at UV wavelengths.  Examples of
flickering can be seen in the residuals of the individual eclipse
observations with respect to the average light curve in the UV first-order
light (see the lowest left panel of Figure~\ref{fig_lcs}).  The typical
flickers last from 1 -- 3 minutes and are 5\% -- 10\% above the local mean.
Some of the brighter flickers reach 25\% -- 40\% above the local mean.

\begin{figure}
\psfig{file=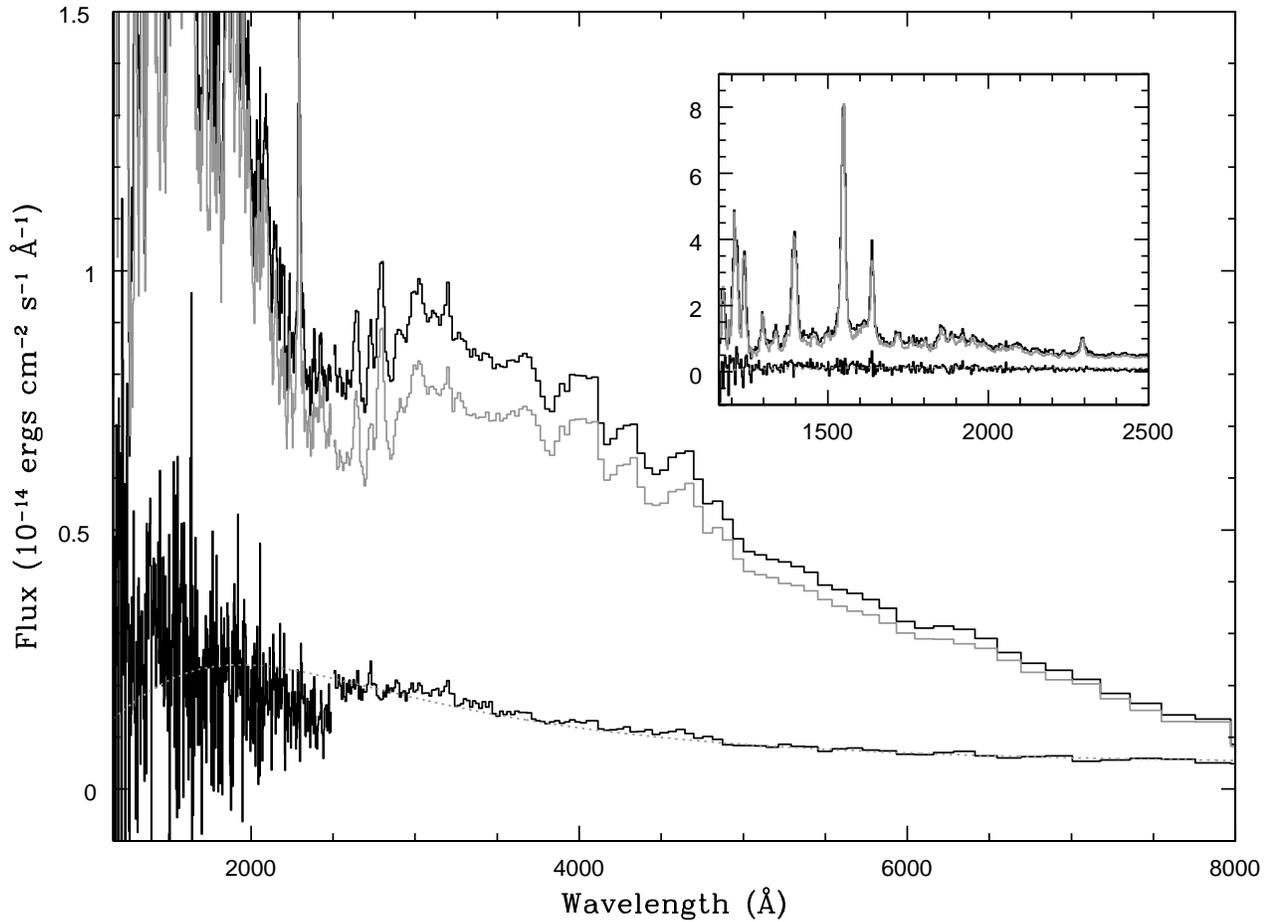,angle=-90,width=7in}
\figcaption[flickersp.ps]{Spectra of V348~Pup taken from the peaks
and troughs of flickers in uneclipsed light.  The upper curve
shows the flickering peaks: mean G160L ($\lambda <1200$~\AA) and
PRISM ($\lambda > 1200$~\AA) spectra of individual spectra more
than 1$\sigma$ above the normalized uneclipsed flux.  The middle
curve, in gray, shows the flickering troughs: mean G160L and PRISM
spectra for individual spectra more than 1$\sigma$ below the
normalized uneclipsed flux.  The bottom curve shows the difference
spectrum.  Also shown for reference is an arbitrarily scaled
blackbody with a temperature of 15,000~K.  The spectra are plotted
in units of $F_{\nu}$ (mJy) vs.\ $\lambda$ (\AA).
\label{fig_flicker}}
\end{figure}

To examine the spectrum of the flickering source, we created an average of
all the spectra obtained at flux peaks and another of the troughs. We
normalized the individual observation light curves, using a constant
normalization for the regions out of eclipse on the eclipse observations
and a low-order (4th order) spline on the two post-eclipse observations.
We then defined the flickering peaks for each observation as comprising all
out of eclipse spectra more than 1$\sigma$ above the mean.  Similarly, the
flickering troughs comprise all spectra more than 1$\sigma$ below the mean.
The resulting spectra and the difference spectrum are shown in
Figure~\ref{fig_flicker}.  The difference spectrum between the peaks and
troughs gives the spectrum of the flickering.  It is featureless,
indicating that the instantaneous flickering occurs in the continuum and
not in the emission lines.  An arbitrarily-scaled, 15,000~K blackbody is
shown in comparison to the flickering spectrum in
Figure~\ref{fig_flicker}. While the blackbody spectrum is not a fit, it
does provide a good match to the shape of the flickering spectrum.

We also examined auto-correlation and cross-correlation functions of light
curves in several passbands to determine if there are any periodicities in
the light curves or a delayed response to the continuum flickering in the
emission lines.  For this, we used the two out of eclipse observations
(Runs 5 \& 6). We created post-eclipse light curves of continuum regions
and emission lines by averaging the flux in each passband and each
spectrum. Emission line light curves in the G160L observation were made for
\ion{Si}{4}, \ion{C}{4}, \ion{He}{2}, and \ion{C}{3} $\lambda$2300~\AA, as
well as for the blend of weaker lines from 1833 -- 2018~\AA.  Continuum
light curves were made for 1429 -- 1511~\AA, 1679 -- 1700~\AA, 2343 --
2493~\AA.  From the PRISM observation, we made emission line light curves
for \ion{C}{3} and \ion{Mg}{2}, and continuum light curves from 1720
--1835~\AA, 2000 -- 2275~\AA, 2340 -- 2620~\AA, 2865 -- 3165~\AA, and 3240
-- 3595~\AA.  We did not make narrow-band light curves at the longer
wavelengths because the variability drops off longward of the Balmer limit.
Finally, we made light curves of the overall spectral regions covered by
each observation, 1150 -- 2500~\AA\ for the G160L and 2000 -- 8000~\AA\ for
the PRISM.

Both post-eclipse observations showed a steady decrease in flux, so we
smoothed the light curves with a high-pass filter (71 pixel boxcar) to
remove the slow variability before conducting the power spectrum analysis.
For this, we used a 71 pixel boxcar filter; each pixel in the light curve
covers 4.29~sec, so this corresponds to filtering over a 5.1~min moving
window (the total observation time was 38~min for each
observation). Experiments with filters of other sizes altered the S/N of
the correlation functions but did not change our results. We also
subtracted the nearby continuum light curve from the line light curves to
focus on the line variability only.

None of the auto-correlation functions of the continuum or line light
curves showed any lagged correlations in the variability.  The continuum
light curves were cross-correlated at zero lag, but there was no
significant lagged correlation between different continuum regions; nor was
there any correlation between continuum and lines at zero or non-zero
lag. Power spectrum analyses of the light curves confirmed this result,
failing to show any significant periodicities in the light curves.  By
injecting sinusoids of known period and amplitude into our broadband light
curves, we determined that we would certainly (probably) have detected
periodicities with an amplitude of 5\% (3\%) in a frequency range of 0.002
-- 0.1~Hz.

\subsection{Model Accretion Disk Fits to the Spectra} \label{sec_disk}

We generated a grid of steady-state accretion disk model spectra
to compare to the UV-optical spectrum of V348~Pup.  The disk
spectra were constructed from summed, area-weighted,
Doppler-broadened model atmosphere stellar spectra.  The stellar
spectra were used to model each accretion disk annulus, with each
spectrum chosen to correspond to the theoretical, steady-state
accretion disk temperature and gravity of that annulus.  For a
detailed description of this procedure, see \citet{long1994}.  The
stellar spectra were created using the model atmosphere and
spectral synthesis codes of Hubeny, TLUSTY and SYNSPEC
\citep{hubeny1988,hubeny1994}.  For temperatures below 12,000~K,
we used the ATLAS model atmospheres of \citet{kurucz1993}. The
stellar atmosphere grid spans a temperature range of 4000 --
100,000~K and $\log g$ of 2.0 -- 8.5. From these, we created model
accretion disk spectra with accretion rates of $\log \dot{m}$ =
15.0 -- 19.0~g~s$^{-1}$ in increments in the log of 0.2.

For the geometry of the system, we adopted the parameters of
\citet{rolfe2000} and \citet{rodriguez2001}, who agree on the
values for which they overlap.  We set $q$ = 0.31, $M_{WD} = 0.65
\msun$, $i = 81\arcdeg$, and $P_{orb}$ = 2.44~hr.  We set $R_{WD}=
\expu{8.5}{8}{cm}$, based on a standard WD mass-radius relation
\citep{nauenberg1972}. The distance to V348~Pup is unknown, so we
fixed the distance in the models at 100~pc.  The normalization of
the models to the observed flux will give a distance for a given
model and accretion rate, but without a distance estimate we do
not have independent information on the best fit model from the
spectral shape and the flux. There are no previous estimates of
the interstellar reddening along the line of sight to V348~Pup, so
the reddening E(B--V) was left as a free parameter in the fits.
There is evidence that the reddening to V348~Pup is not very high:
a dust map of the sky (with a resolution of $6\farcm1$) indicates
that the total extinction looking out of the Galaxy in the
direction of V348~Pup is E(B--V) = 0.216 (Schlegel et al.\ 1998;
calculated using the NASA/IPAC Extragalactic Database).

The accretion disk models are circular, not elliptical, which differs from
the actual geometry of the disk in V348~Pup, at least in its outermost
radii \citep{rolfe2000}. Moreover, \citet{rolfe2000} show that the size and
the ellipticity of the disk varies over the superhump cycle. As a result,
we tested two disk sizes in our models.  In the first, we used a disk that
has a large inner radius ($R_{inner} = 0.1 R_{L_{1}}$) and an outer radius
that extends to the tidal cut-off radius, $R_{outer} = 0.9 R_{L_{1}}$, to
mimic the Rolfe et al.\ disk parameters. Second, we used a more
``standard'' disk whose inner edge reaches to the WD surface ($R_{inner} =
0.03 R_{L_{1}}$) and that has a smaller outer radius, $R_{outer} = 0.4
R_{L_{1}}$, the circularization radius of the disk \citep{frank1992}.  We
fit the models by least squares to a concatenation of the grating and prism
out-of-eclipse spectra (switching between settings at 2500~\AA), and
excluded strong emission lines from the fit.

\begin{figure}
\psfig{file=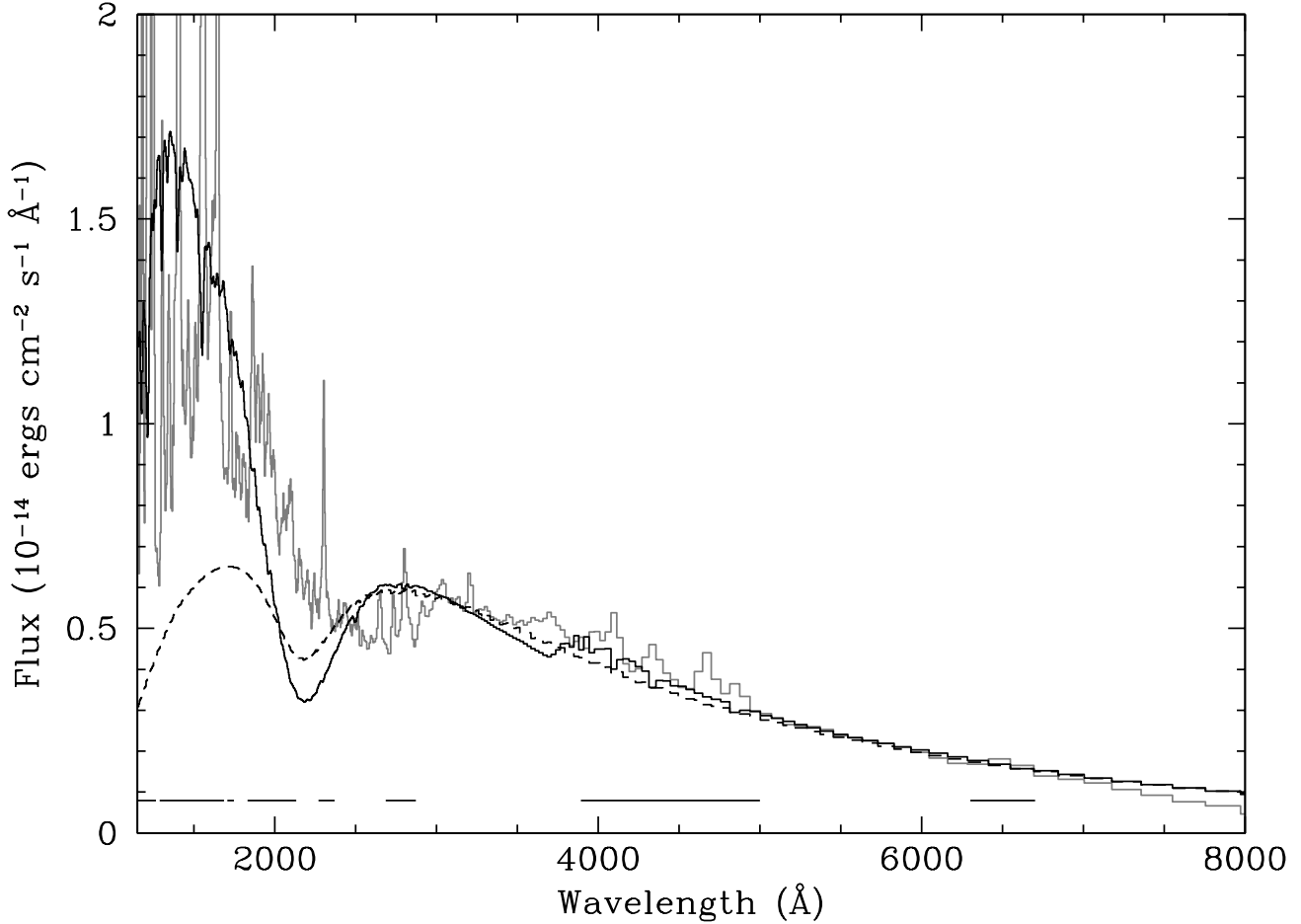,angle=-90,width=7in}
\figcaption[diskbest.ps]{The best fit accretion disk model compared to the
out of eclipse G160L and PRISM spectra. The observed spectrum, shown as the
gray line, is a concatenation of the G160L and PRISM out of eclipse
spectra.  The transition between the two settings occurs at 2500~\AA.
Shown as the black solid line is the best-fit accretion disk model assuming
the Rolfe et al.\ disk. The model parameters give a mass accretion rate of
$\dot{m} = \expu{1}{19}{g \: s^{-1}}$ and a reddening of E(B--V) = 0.40.
The normalization of the model to the data is N = 0.04, which gives a
distance of 500~pc.  The reduced chi-squared of the fit is $\chi^{2}_{\nu}
= 310$. Also shown as the dashed line is the best-fit accretion disk model
when the disk is modeled with sums of blackbodies rather than with sums of
stellar spectra. The model parameters are similar to those in the first
model: $\dot{m} = \expu{6.3}{17}{g \: s^{-1}}$, E(B--V) = 0.18, and N =
0.057 (429~pc).  The fit gives $\chi^{2}_{\nu} = 212$. The solid bars at
the bottom of the figure show regions that were masked out during the
fitting.
\label{fig_diskbest}}
\end{figure}

Figure~\ref{fig_diskbest} shows the best fit accretion disk model when the
Rolfe et al.\ disk size is used. The model has a mass accretion rate of
$\dot{m} = \expu{1}{19}{g \: s^{-1}}$, a reddening E(B--V) = 0.40, and a
normalization N = 0.04, which gives a distance of 500~pc for a disk with
this flux and accretion rate.  The model fit to the data is poor both
qualitatively and statistically ($\chi^{2}_{\nu} = 310$).  The maximum
allowed accretion rate is selected, primarily to weaken the amplitude of
the Balmer jump, although it remains larger in the model than in the
observed spectrum.  Despite the large reddening, the model still exceeds
the observed UV flux. Similar problems were encountered with fits to the
UV-optical spectrum of UX~UMa \citep{knigge1998}. The reddening value also
results in a prominent 2175~\AA\ feature.  As can be seen, the model
2175~\AA\ feature is much deeper than the observed spectrum at that
wavelength.  It is also clear that interstellar reddening cannot explain
the 2000 -- 3000~\AA\ dip in the observed spectrum, which is broader and
comes to a minimum at a longer wavelength (2600~\AA) than the 2175~\AA\
reddening feature.

A change to the smaller accretion disk size fails to improve the
fit.  The fit is qualitatively similar to the previous fit and
$\chi^{2}_{\nu} = 315$.  The disk parameters are not the same,
however: $\dot{m} = \expu{2.5}{18}{g \: s^{-1}}$, E(B--V) = 0.50,
and N = 0.16 (251~pc).  The mass accretion rate drops and the
reddening increases compared to the previous model because the
``standard'' disk extends to the surface of the WD, which results
in more disk emission in the blue.  The UV emission still exceeds
the observed UV continuum flux and the resulting 2175~\AA\ feature
dips 60\% below the observed flux at that wavelength.

In order to distinguish between possible problems with the
assumption of a steady-state disk dominating the continuum and
problems with the use of stellar spectra to model the disk, we
also show in Figure~\ref{fig_diskbest} the best-fit accretion disk
spectrum when the disk is modeled as a sum of blackbodies at each
annulus rather than stellar spectra.  The model parameters are
$\dot{m} = \expu{6.3}{17}{g \: s^{-1}}$, E(B--V) = 0.18, and N =
0.06 (419~pc).  The fit to the spectrum is improved
($\chi^{2}_{\nu}$ = 212), mainly because of a lack of a Balmer
jump and less UV emission in the blackbody model.  The accretion
rate and reddening are also at lower, less extreme values.
Nevertheless, the fit to the observed spectrum remains poor. The
reddening dip at 2175~\AA\ is still too large, and the model
overshoots the observed flux around 2500 -- 3500~\AA. In this
case, the UV flux is too low rather than too high compared to the
observed spectrum.

\begin{figure}
\psfig{file=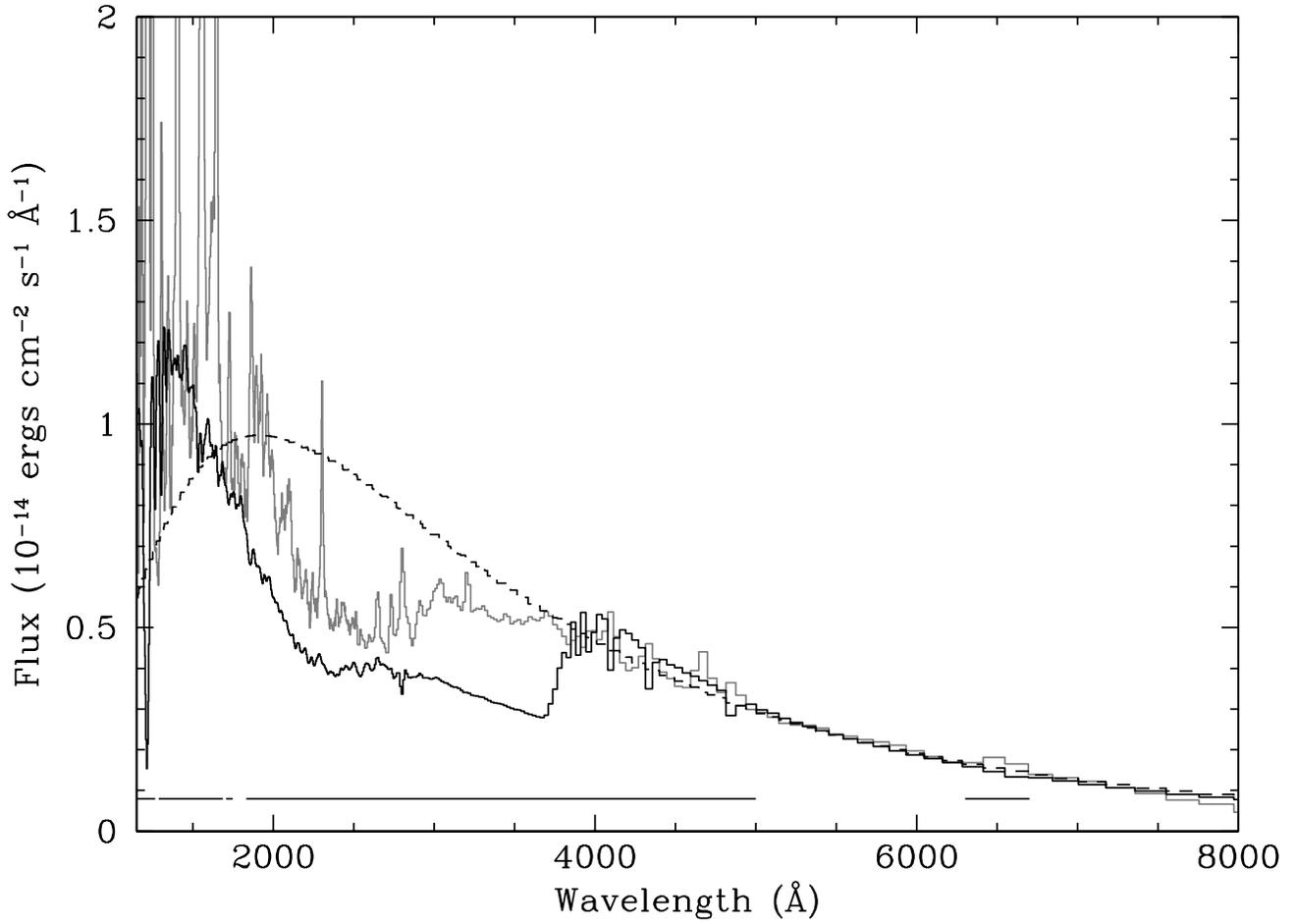,angle=-90,width=7in}
\figcaption[disknext.ps]{The best fit accretion disk models when the 2000 --
3000~\AA\ dip and the Balmer jump are excluded from the fit regions.  The
model parameters are: $\dot{m} = \expu{5.2}{17}{g \: s^{-1}}$, E(B--V) =
0.11, and N = 0.058 (415~pc) for the model constructed as sums of stellar
spectra and $\dot{m} = \expu{3.6}{17}{g \: s^{-1}}$, E(B--V) = 0.002, and N
= 0.048 (4561~pc) for the model constructed from sums of blackbodies. The
fits give $\chi^{2}_{\nu} = 70$ and $\chi^{2}_{\nu} = 80$,
respectively. \label{fig_disknext}}
\end{figure}

Since the observed spectral dip and lack of a Balmer jump are what drove
the model disk fits to their extreme parameter values, we also tested fits
to the spectrum in which we masked out all wavelengths between 2000 and
5000~\AA.  The best fit in this case (again, for the Rolfe et al.-like
disk) is shown in Figure~\ref{fig_disknext}.  The model parameters are
$\dot{m} = \expu{5.2}{17}{g \: s^{-1}}$, E(B--V) = 0.11, and a
normalization N = 0.06, giving a distance of 415~pc. The $\chi^{2}_{\nu}$
is 70.  Outside of the Balmer jump region, this fit is better than the
previous model fits.  The UV flux is more consistent with the observed
level, and the lower reddening gives a near-UV slope and weak 2175~\AA\
feature that is also closer to that observed.  Unfortunately, this model
also has a Balmer jump in which the flux drops by a factor of two, when no
Balmer jump is observed.  When the ``standard'', disk is used, the fit is
qualitatively similar, but the parameters are $\dot{m} = \expu{8.1}{15}{g
\: s^{-1}}$, E(B--V) = 0.001, and a normalization N = 0.39
(160~pc). $\chi^{2}_{\nu}$ is 59.  As a result, even if we are inclined to
ignore problems with the Balmer jump and accept these models, we cannot
constrain the mass accretion rate to better than a factor of 64 without
knowing more about the structure of the disk or the distance to the system.

Finally, we also show in Figure~\ref{fig_disknext} the model fit for the
``standard disk'' when the disk is constructed from sums of
blackbodies. The model parameters are $\dot{m} = \expu{3.6}{17}{g \:
s^{-1}}$, E(B--V) = 0.002, and N = 0.05 (456~pc). $\chi^{2}_{\nu} =
80$. The mass accretion rate and normalization are similar to those of the
disk model constructed from sums of stellar spectra, but the smaller UV
flux of the blackbody model results in a lower, effectively zero,
reddening.  The model is a reasonable fit to the UV and red optical fluxes,
and there is by definition no Balmer jump problem, but the model overshoots
the observed spectrum in the region of the dip, 2000 -- 3500~\AA.

\section{Discussion} \label{sec_discussion}

One notable characteristic of the depth of the eclipse in V348~Pup is its
relative flatness over the UV-optical range.  The depth of the continuum
eclipse at 1300~\AA\ is the same as the eclipse depth at 5500~\AA, and the
fraction of the out of eclipse flux occulted only drops slightly
($\simeq$5\%) at longer wavelengths.  Because NL CVs do not show outbursts,
their accretion flow is assumed to be largely steady (leaving aside
intermittent low states, which do not concern us here).  The theoretical
radial temperature profile of a steady-state disk shows a decline in disk
temperature with increasing disk radius, with $T(R) \propto R^{-3/4}$
\citep{pringle1981}. The high viewing inclination of V348~Pup ($i =
81\arcdeg$) ensures that the central plane of the inner disk is fully
occulted at mid-eclipse. Therefore, the eclipse of a steady-state disk
should be deeper at shorter wavelengths as the hotter parts of the disk are
fully occulted, while the cooler, outer disk is uneclipsed.  This is not
the case in the eclipse of V348~Pup.

The shape of the eclipse light curves bolsters this point.  The
light curves in V348~Pup are largely V-shaped, particularly in
contrast to the smooth, U-shaped eclipses of UX~UMa. V-shaped
eclipses are indicative of the eclipse of a disk with a flat
temperature distribution; this deviation from the expected
steady-state temperature profile has been seen in other NLs of the
SW~Sex subclass (e.g., Rutten et al. 1992).  Finally, the poor
fits of theoretical steady-state accretion disk models to the
time-averaged spectrum of V348~Pup indicate that the continuum is
not well described by this model.  Admittedly, there have been
problems fitting disk models created from sums of stellar spectra
in other systems in which the continuum is believed to be
dominated by steady-state accretion disk emission (such as
UX~UMa), but in our modeling, even theoretical spectra created
from sums of blackbodies failed to match the overall shape of the
UV-optical continuum.

There are a couple of possibilities for why the continuum shape and eclipse
are not consistent with the expected properties of a steady-state disk. The
first is that the disk is not in a steady-state. \citet{rutten1992} offer
several possibilities for the flat T(R) distribution is the SW~Sex-type NLs
they modeled, including disruption of the inner disk by the WD magnetic
field, an extended boundary layer, or a substantial outflow. Another
possibility, for which there is empirical evidence in a similar system, is
that we are not viewing the inner disk because it is shielded. The NL
DW~UMa is a SW~Sex star with the same inclination as V348~Pup. When DW~UMa
went into an optical low state, its UV flux actually increased
\citep{knigge2000}. Its normal UV spectrum is very similar to the UV
spectrum of V348~Pup, while the UV spectrum in the optical low state was
that of a WD. This led Knigge et al.\ to conclude that the WD and the inner
disk are usually invisible in DW~UMa, blocked by a self-occulting, flared
accretion disk.

Because of the similarity in binary parameters, inclination, and
spectral morphology between DW~UMa and V348~Pup, this is an
attractive scenario to explain the flat T(R) profile in V348~Pup
as well. Indeed, \citet{knigge2000} suggest that a self-shielding
disk could be a defining characteristic of the SW~Sex stars, most
of which are eclipsing (although it should be reiterated that it
is not clear if the preponderance of eclipsing systems among the
SW~Sex population is a characteristic of the class or is a
selection effect).  In V348~Pup, the absence of any WD ingress
features in the light curves and the extreme UV excess (even with
implausibly high reddening values) in the model disk fits support
the idea that the inner disk and the white dwarf are shielded from
view.

It is also clear that much of the material we see being eclipsed
lies above the disk plane. Because of its high inclination, the
inner annuli --- $\leq0.42R_{L_{1}}$, assuming $q$ and $i$ from
Rolfe et al.\ 2000 and Rodr\'{\i}guez-Gil 2001 --- and the back of
the accretion disk are fully occulted in V348~Pup. However, it is
the plane of the disk that is occulted; vertically extended
material ($z/R \geq 0.16$), even at the center of the disk, will
not be occulted at mid-eclipse. All of the emission lines show
more shallow eclipses than those of the adjacent continuum,
indicating that the vertical extent of the lines exceeds that of
the continuum.  The strongest UV lines, in particular, are only
weakly eclipsed. Shallow or non-existent eclipses are a common
characteristic of UV emission lines in CVs
\citep{holm1982,cordova1985,drew1985}.  The extended line emission
may originate in wind outflows and/or a vertically extended disk
chromosphere. It is notable that the asymmetry of the UV lines in
eclipsed light is seen in the strongest lines in the spectrum,
which also show the weakest eclipses. These lines likely have the
largest vertical extension of line emission above the disk, such
that we continue to see scattered red-shifted emission from the
outflowing gas even as the parts of the line nearest the disk are
occulted.

The broad spectral dip extending from 2000 -- 3000~\AA\ is, like
the line emission, eclipsed less than the continuum, suggesting
that the source of the dip is also vertically extended with
respect to the accretion disk continuum. The dip is similar to a
feature seen in the spectrum of the eclipsing dwarf nova, OY~Car.
\citet{horne1994} successfully modeled the dip in OY~Car as
absorption of the underlying continuum by an ``\ion{Fe}{2}
curtain'' of vertically extended, veiling gas. The 2000 --
3000~\AA\ dip is not seen in the FOS spectrum of the
lower-inclination system UX~UMa, but it appears to be present in
DW~UMa. The continuum V348~Pup is also not as blue as in UX~UMa,
although the emission lines are stronger.  All of these
differences are consistent with inclination effects.  The redder
continuum in V348~Pup is probably a result of greater shielding of
the inner disk in the higher-inclination system.

There is little in our observations to distinguish V348~Pup, a
rare disk-accreting NL CV found in the period gap, from longer
period NL CVs. It has been proposed that V348~Pup is a magnetic CV
\citep{tuohy1990}.  If true, then its orbital period would place
it near the important phase of synchronization of the WD rotation
and orbital periods in some evolutionary models of magnetic CVs
\citep{webbink2002}.  We find no evidence in these observations to
support the V348~Pup as a magnetic CV, however. In particular, we
did not find any periodic variability that would signal the
presence of a magnetic WD (although it is certainly possible that
a self-shielding disk could hide the signature of a magnetic WD).
Rather, in its UV-optical spectrum and eclipse behavior, V348~Pup
behaves as a typical SW~Sex system.  Unfortunately, the poor model
accretion disk fits to the observed spectrum preclude a
determination of the mass accretion rate in V348~Pup, which may
have indicated whether there are differences between luminous
period gap CVs and their brethren outside of the gap.

\section{Conclusions} \label{sec_conc}

We have obtained high time series, moderate to low spectral
resolution, UV and optical spectrophotometry of the eclipsing NL
V348~Pup.  Three of the observations in each setting were obtained
through eclipse while one observation in each setting covered the
orbital phases after the eclipse. The results of the observations
are as follows:

\begin{enumerate}

\item The time-averaged spectrum outside of eclipse shows a steady rise in
flux to shorter wavelengths, although there is a broad spectral dip
extending from 2000 -- 3000~\AA. Superposed on the continuum are strong
emission lines of \ion{H}{1}, \ion{He}{2}, and resonance and excited state
transitions of ionized metals.  There were no gross changes in the shape or
flux of the spectrum over the 6 days of our observations.

\item The continuum eclipse is relatively flat over the UV-optical
spectral range; 55\% -- 60\% of the out of eclipse continuum flux
is occulted at mid-eclipse at most wavelengths.  The eclipse is
more shallow in the 2000 -- 3000~\AA\ trough, where 45\% -- 50\%
of flux is eclipsed. The eclipse light curves in the continuum are
wide and V-shaped. There is little change in eclipse shape with
wavelength, although the eclipses at the reddest wavelengths are
slightly broader and shallower than the eclipses in the blue and
UV. The eclipses are largely featureless; in particular, there is
no sign of the eclipse of the WD in any of the light curves.

\item All of the emission lines remain present in eclipse.  All of the
lines show a more shallow eclipse than the nearby continuum.  The strongest
UV lines show the least amount of occultation during eclipse. For most of
the emission lines, the eclipse is symmetric about the line centers and the
line profile does not change in eclipse, but in \ion{C}{4}
$\lambda$1550~\AA\ the eclipse is skewed to wavelengths blue of the line
peak, while the red component of the line becomes stronger at mid-eclipse.
There is also a mild asymmetry toward a weaker eclipse at red velocities in
\ion{C}{3} $\lambda$1175~\AA\ and $\lambda$1855~\AA, and \ion{Si}{4}
$\lambda\lambda$1393,1402~\AA.

\item Spectra pre- and post-eclipse show that most of the emission lines
and the blue continuum are weaker after eclipse than before. The dip in the
spectrum from 2000 -- 3000~\AA\ is also deeper after eclipse than before.

\item Flickering is strong in the observations. Typical flickers
last from 1 -- 3 minutes, and range from 5\% -- 15\% in excess of
the local flux levels. A flickering spectrum shows that flickering
occurs only in the continuum and that the flickering source is
qualitatively well fit by a 15,000~K blackbody spectrum.  We did
not detect any periodicites associated with the variability.

\item Model steady-state accretion disk spectra give poor fits to
the shape of the UV-optical spectrum.  Disk spectra constructed
from sums of stellar spectra are too strong in the UV and have
Balmer jumps well in excess of the observed feature. Disk spectra
constructed from sums of blackbodies do not have these problems,
but they are still poor fits to the shape of the observed
spectrum.  The 2000 -- 3000~\AA\ dip, in particular, is not fit by
the model disk spectra.  Due to its width and the wavelength of
its minimum, this feature is not attributable to reddening.

\item The UV-optical properties of V348~Pup near and through eclipse are
consistent with the behavior of SW~Sex type NL CVs, and support the
identification from \citet{rodriguez2001} of V348~Pup as a member of this
class.  The high inclination, shallow and flat eclipse, and similarity to
the spectrum of DW~UMa suggest that the accretion disk is self-occulting,
blocking a line of sight view of the inner disk and the WD. The broad
spectral dip is consistent with the presence of a vertically-extended,
absorbing ``\ion{Fe}{2} curtain'' in the system. The UV-optical eclipse
observations do not reveal any characteristics of the system to explain its
location in the period gap.  There are no periodicities in the light curves
to indicate the presence of a magnetic WD.

\end{enumerate}

\acknowledgements

These observations are associated with proposal GO--6796. Support
for proposal GO--6796 was provided by NASA through a grant from
the Space Telescope Science Institute, which is operated by the
Association of Universities for Research in Astronomy, Inc., under
NASA contract NAS 5-26555.  We gratefully acknowledge the support
from NASA for this project. RB acknowledges financial support from
CNPq/Brazil through grant no.\ 300354/96-7.


\pagebreak

\newpage
\pagestyle{empty}

\newpage
\pagestyle{empty}

\newpage
\pagestyle{empty}

\newpage
\pagestyle{empty}

\newpage
\pagestyle{empty}

\newpage
\pagestyle{empty}

\newpage
\pagestyle{empty}

\newpage
\pagestyle{empty}

\newpage
\pagestyle{empty}

\newpage
\pagestyle{empty}

\newpage
\pagestyle{empty}

\clearpage
\begin{deluxetable}{lcccccc}
\tablecaption{Observation Summary\label{tab_obs}}
\tablewidth{0pt}
\tablecolumns{7}
\tablehead{
\colhead{Run} & \colhead{Date} & \colhead{Start} &
\colhead{End} & \colhead{Phase Range\tablenotemark{a}} &
\colhead{Spectra Obtained\tablenotemark{b}} & \colhead{Detector} \\
\colhead{} & \colhead{(1996)} & \colhead{(UT)} &
\colhead{(UT)} & \colhead{(Cycles)} &
\colhead{} & \colhead{} }
\startdata
1\dotfill & Sept 11 & 19:11:08 & 19:51:23 & 0.91 -- 1.20 & 578 & PRISM \\
2\dotfill & Sept 14 & 18:04:49 & 18:43:30 & 0.91 -- 1.18 & 542 & G160L \\
3\dotfill & Sept 14 & 22:55:50 & 23:23:27 & 0.90 -- 1.16 & 527 & PRISM \\
4\dotfill & Sept 15 & 03:44:22 & 04:23:03 & 0.86 -- 1.13 & 542 & G160L \\
5\dotfill & Sept 15 & 21:28:24 & 22:06:01 & 0.12 -- 0.38 & 527 & PRISM \\
6\dotfill & Sept 16 & 02:17:04 & 02:55:45 & 0.09 -- 0.36 & 542 & G160L \\
7\dotfill & Sept 17 & 12:06:38 & 12:44:15 & 0.93 -- 1.19 & 527 & PRISM \\
8\dotfill & Sept 17 & 21:44:26 & 21:23:07 & 0.87 -- 1.14 & 542 & G160L \\
\enddata
\tablenotetext{a}{Orbital phasing based on the photometric
ephemeris of Baptista et al.\ (1996). We have shifted the absolute
phasing by 0.024 cycles to bring the eclipses symmetric about
phase 0.} \tablenotetext{b}{Each spectrum has an exposure time of
4.28~sec, with 0.1~sec of dead time between exposures.}
\end{deluxetable}

\end{document}